\documentclass[runningheads]{llncs}
\usepackage{cite}
\usepackage{amsmath,amssymb,amsfonts}
\usepackage{algorithmic}
\usepackage{algorithm}
\usepackage{graphicx}
\usepackage{textcomp}
\usepackage{epstopdf}
\usepackage{multirow}
\usepackage{booktabs}
\usepackage[table,xcdraw]{xcolor}
\usepackage{float} 
\usepackage{subfigure} 
\usepackage[misc]{ifsym}
\usepackage{hyperref}
\usepackage{pdfpages}

\begin{document}
\title{Adversarial Learning Based Structural Brain-network Generative Model for Analyzing Mild Cognitive Impairment}
%
%
\author{Heng Kong\inst{1, 2} \and
Shuqiang Wang\inst{2}}

\institute{Southern University of Science and Technology, Shenzhen, 518000, China \\
\email{12132527@mail.sustech.edu.cn}\\ \and
Shenzhen Institutes of Advanced Technology, Chinese Academy of Sciences, Shenzhen, 518000, China \\
\email{\{sq.wang\}@siat.ac.cn}
}
\maketitle              
\begin{abstract}
	Mild cognitive impairment(MCI) is a precursor of Alzheimer's disease(AD), and the detection of MCI is of great clinical significance. Analyzing the structural brain networks of patients is vital for the recognition of MCI. However, the current studies on structural brain networks are totally dependent on specific toolboxes, which is time-consuming and subjective. Few tools can obtain the structural brain networks from brain diffusion tensor images. In this work, an adversarial learning-based structural brain-network generative model(SBGM) is proposed to directly learn the structural connections from brain diffusion tensor images. By analyzing the differences in structural brain networks across subjects, we found that the structural brain networks of subjects showed a consistent trend from elderly normal controls(NC) to early mild cognitive impairment(EMCI) to late mild cognitive impairment(LMCI): structural connectivity progressed in a progressively weaker direction as the condition worsened. In addition, our proposed model tri-classifies EMCI, LMCI, and NC subjects, achieving a classification accuracy of 83.33\% on the Alzheimer's Disease Neuroimaging Initiative(ADNI) database.

\end{abstract}
\section{Introduction}
Alzheimer's disease(AD) is an irreversible neurodegenerative disease with a high prevalence in the elderly population, which is characterized by a severe clinical decline in memory and cognitive functions\cite{ad1}. The pathogenic mechanisms of AD are not fully understood\cite{ad2}, and there is a lack of effective therapeutic drugs\cite{ad3}\cite{ad4}. Mild cognitive impairment(MCI), a prodromal stage of AD\cite{mci2}\cite{mci3}\cite{mci4}, has attracted widespread interest in the study of patients with MCI because interventions at this stage can reduce the rate of conversion to AD. In order to detect AD patients as early as possible, medical image computational methods are used to help diagnose this type of brain disease with the help of computer vision\cite{noganad1}\cite{noganad2}. As a critical method in the artificial intelligence field, deep learning has been widely applied in medical image computing\cite{nogannoad1}\cite{nogannoad2}. Because of its excellent performance, many complex patterns of medical imaging can be identified, and it enables the auxiliary diagnosis of disease\cite{ganad3}\cite{noganad3}. Medical images of AD have their complicated pattern and biomarkers, and a deep learning-based approach can handle them through its specific techniques. For example, convolution neural networks(CNN) is a universal imaging computing method. It can automatically learn the parameters of the convolution kernel and extract features that are highly correlated with the target\cite{cnn3}\cite{cnn4}. Notably, 3D CNN has better processing performance for 3D brain image data such as DTI. However, the patterns of high-dimensional medical images are difficult to analyze and identify. Brain networks as a pattern of networks integrated by connections can better and more easily represent downscaled brain information. They can be derived from anatomical or physiological images, yielding structural and functional networks\cite{conection2}\cite{conection3}, respectively. Structural connectivity of brain networks describes the anatomical connections that connect a set of neuronal elements. Diffusion tensor imaging(DTI) shows changes in the number of fiber bundle connections, fiber bundle length, and anisotropy values, which can provide valid structural information for MCI detection\cite{cnn5}.

The data of medical images have problems such as a small sample size and uneven distribution of categories. These problems are prone to overfitting and other problems when applying traditional deep learning methods. The generative adversarial network(GAN) method is more suitable for the medical image field due to the use of the idea of adversarial learning. GAN can be seen as variational-inference\cite{bianfen1} based generative model. Recently, more and more researchers have used the GAN method to study Alzheimer's disease. Especially in the field of brain medical images, many classic GAN methods\cite{ganad4} \cite{ganad5} have emerged. Yu et al.\cite{ganad1} proposed a GAN method that applied tensor decomposition to model training, effectively improving the model's generation effect. Hu et al.\cite{ganad2} proposed a bidirectional mapping mechanism applied to the training of the GAN network, and the generated image is more in line with the real effect. Hu et al.\cite{ganad6} proposed a cross-modal adversarial generation method applied to the generation of brain medical images, effectively improving the model's classification effect. You et al.\cite{ganad7} proposed the FP-GAN model to improve brain medical images' texture generation effectively. Yu et al.\cite{ganad10} proposed the MP-GAN method to successfully extract rich features from different types of MRI images and effectively improve the generation effect of the model. Zuo\cite{ganad8}, Pan\cite{ganad9} and others introduced graph convolution and hypergraph operation\cite{ganad11} into the generative adversarial network and analyzed the pathological mechanism of Alzheimer's disease in the field of whole-brain topology. All of the above methods have promoted the in-depth exploration of the pathogenesis of Alzheimer's disease to some extent in some fields. Nevertheless, these methods still have much room for optimization. This paper proposed a GAN-based method, and the specific contributions are as follows:

This paper proposed an SBGM framework. This method integrates classification and reconstruction knowledge to realize the direct generation of structural brain networks from DTI. The structural networks generated by the model contain structural connection information and feature information between brain regions. The model proposed in this paper is able to classify three types of subjects: NC, EMCI, and LMCI. More importantly, this model provides a generative approach from DTI to structural brain networks. The experimental results on the ADNI database also prove that our proposed model has high classification accuracy. Moreover, the structural brain networks generated by the model have a consistent pattern of change with the structural brain networks obtained from the software template calculations: we found a gradual weakening of structural brain network connectivity in subjects from NC to EMCI to LMCI.

\section{Methods}
\begin{figure}
    \centering
    \includegraphics[width=1\linewidth]{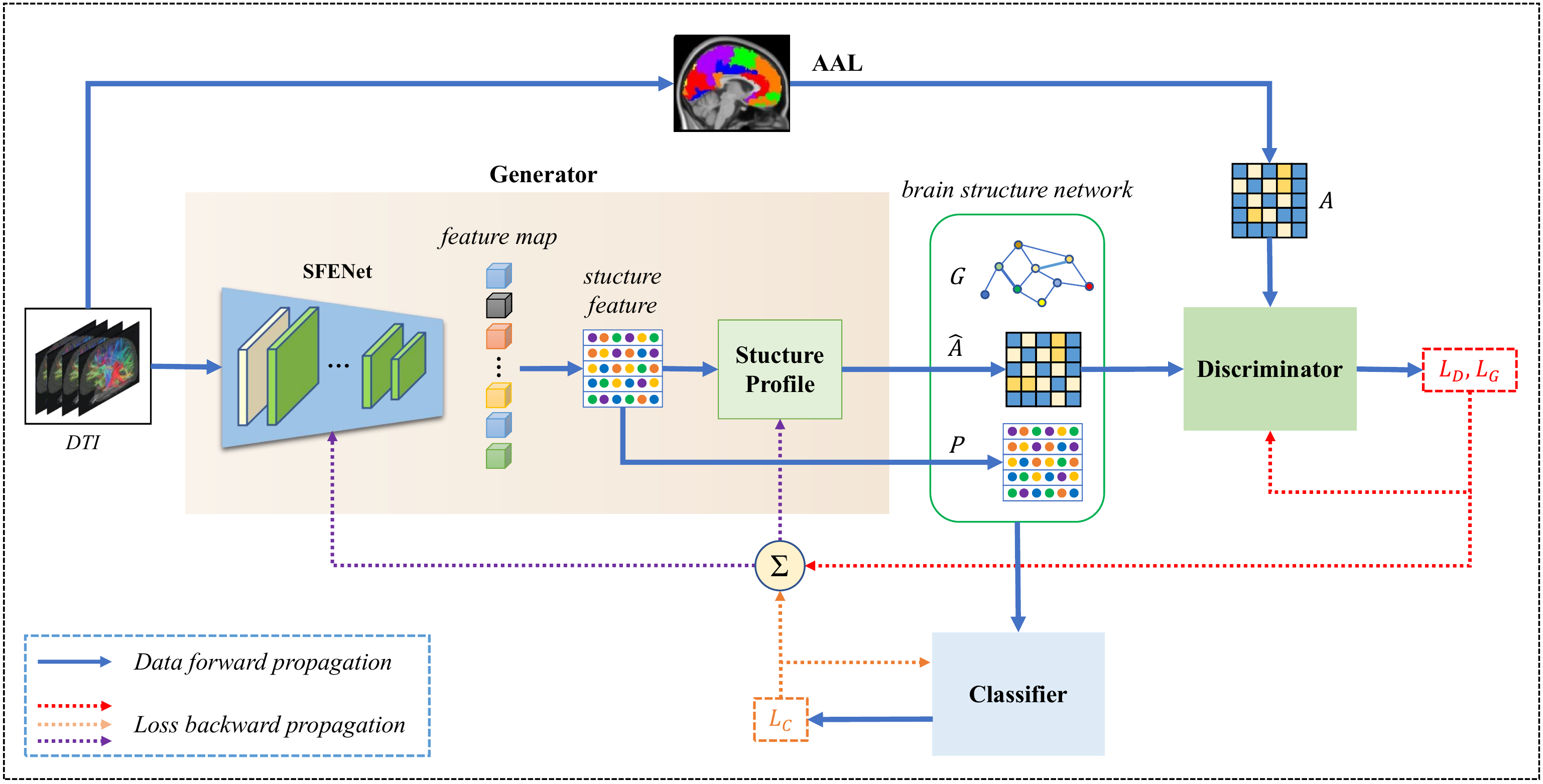}
    \caption{Overall framework of the proposed SBGM.}
    \label{fig:fig1}
\end{figure}

The framework diagram of SBGM is given in Figure \ref{fig:fig1}. Our proposed model consists of three main modules: generator, discriminator, and classifier. The generator consists of two sub-modules, SFENet(Structural Feature Extraction Network) and Structure Profile. SFENet is used to extract structural features of brain regions from 3-dimensional DTI as node features of the structural brain network, and Structure Profile maps structural features of brain regions into a matrix of structural connections between brain regions. The discriminator consists of two fully connected layers to distinguish whether the structural connection matrix comes from the generator or the AAL template. The classifier consists of a graph convolutional network(GCN) and a fully connected layer for the tri-classification of EMCI, LMCI, and NC subjects. The AAL template\cite{AAL1}\cite{AAL2}\cite{AAL3}\cite{AAL4}\cite{AAL5}\cite{AAL6} is a commonly used template for brain region delineation. In this paper, we use it to process the DTI to obtain empirical knowledge of the brain structural connectivity matrix. The meanings of other symbols in the model are as follows: $G$ denotes the topology of the structural brain network, $\hat{A}$ denotes the brain structural connectivity matrix generated by Structure Profile, $P$ denotes the structural features of brain regions extracted by SFENet, $A$ denotes the structural connectivity matrix obtained from the AAL template, $L_D$ denotes discriminator loss, $L_G$ denotes generator loss, and $L_C$ denotes classification loss.

\subsection{Generator}
\paragraph{\textbf{Structural brain network.}}
The generator's main function is to generate the structural network of the brain, which contains the structural connectivity matrix $\hat{A}$ between brain regions and the structural feature \cite{scfeature} matrix $P$. The structural network of brain is a weighted graph, which can be represented as $G(V,E,\hat{A},P)$, where $V=\{v_1,v_2,...,v_N\}$ denotes the $N$ brain regions and $E$ denotes the edges connecting different brain regions\cite{roi}. The structural connectivity matrix $\hat{A}$ represents the relative connection strength between brain regions, $\hat{A}$ is a symmetric matrix, and all elements are continuous values between $[0,1]$. If $\hat{A}_{ij}=\hat{A}_{ji}=0$, indicates that there is no direct structural connection between brain region $v_i$ and $v_j$, then the corresponding edge $E(v_i,v_j)$ does not exist in the structural network of the brain. Generally, the structural connections between most brain regions are weak, so $\hat{A}$ is a sparse matrix. $P$ is the structural feature matrix of brain regions, and each row represents the feature vector $p_i$ of brain region $v_i$. In this study, to make the generated brain structural connectivity matrix have the same dimensionality as the structural connectivity matrix from the AAL template, we take $N=90$ based on empirical knowledge.

\paragraph{\textbf{SFENet.}}
CNN can automatically learn the parameters of the convolution kernel and extract features that are highly relevant to the task\cite{cnn1}\cite{cnn2}, so we use CNN to extract structural features of brain regions from DTI. According to the definition and division of brain regions in neuroscience and medicine, a brain region is a complex irregular chunk in the volume of the brain. In order to fully extract the features of these brain regions, we need to extend the convolution kernel of CNN to 3 dimensions. However, extending the convolution kernel to 3 dimensions increases the parameter size linearly and increases the training difficulty.

\begin{figure}
    \centering
    \includegraphics[width=0.6\linewidth]{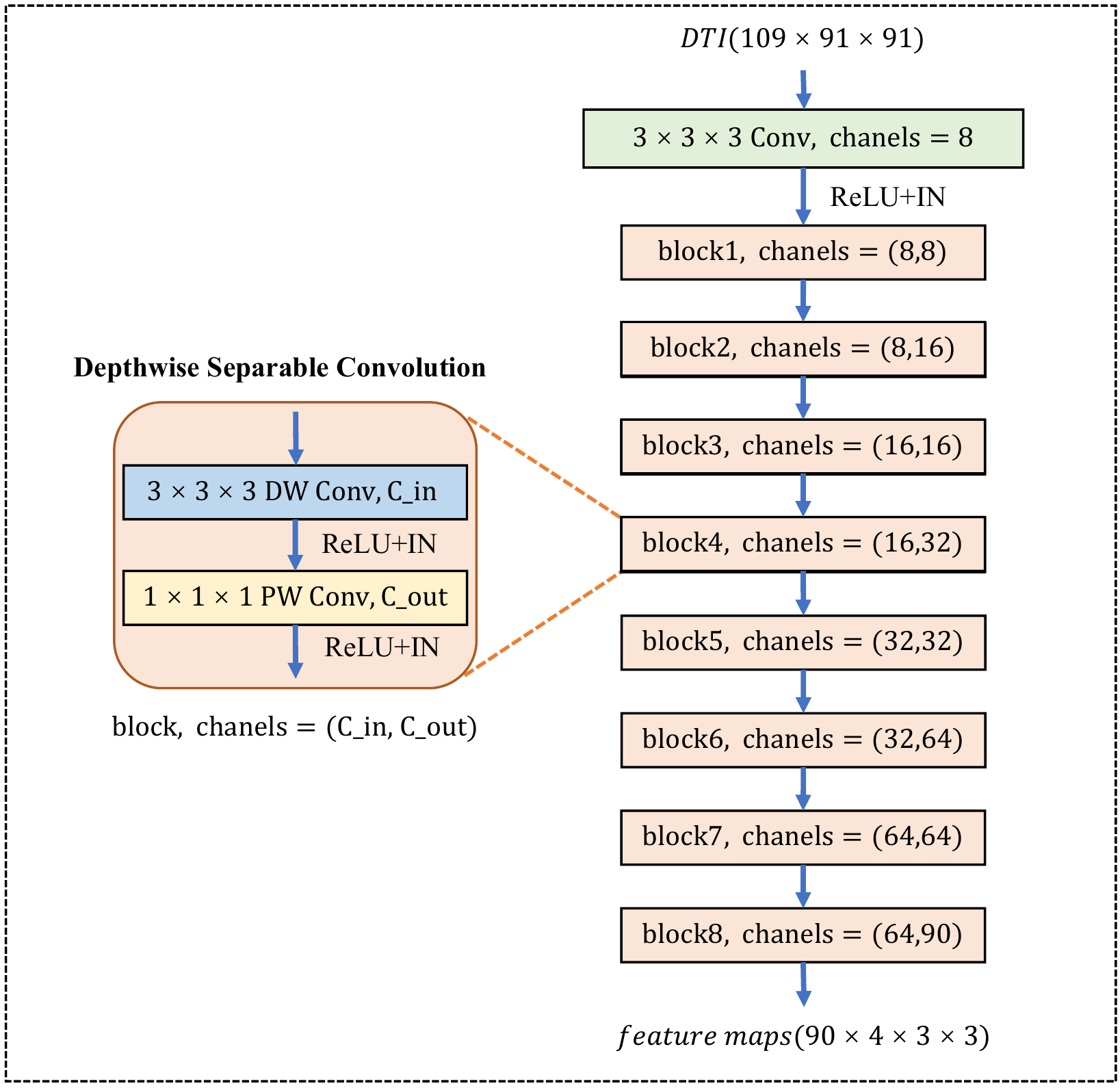}
    \caption{Framework of SFENet.}
    \label{fig:fig2}
\end{figure}

To reduce the training difficulty of the model, we designed a lightweight 3D CNN model-SFENet, as shown in Figure \ref{fig:fig2}. SFENet performs feature extraction for DTI by combining one standard convolutional(Conv) layer and eight depthwise separable convolutional blocks, with each block containing a depthwise convolutional(DW Conv) layer and a pointwise convolutional(PW Conv) layer. The standard convolution layer sets one convolution kernels$(3\times3\times3\times8)$ for each input channel to extract the lower-level structural feature information of DTI. The depthwise convolution layer has one convolution kernel$(3\times3\times3\times1)$ for each input channel, and the number of convolution kernels is the same as the number of channels $C_{in}$ of the input feature map. Depthwise convolution only uses the information of a single input channel, does not take full advantage of the information between channels, and cannot change the number of output channels $C_{out}$. To compensate for the shortcomings of depthwise convolution, we add a pointwise convolution layer afterward to weight the information of all channels and adjust the number of channels of the output feature map. After each convolution, we use a linear collation function(ReLU) as the activation function to alleviate the gradient disappearance problem and perform an instance normalization(IN) operation on the output feature map to prevent “gradient explosion”. The DTI we use is a single channel 3-dimensional image of size $(109\times91\times91)$, and the desired output is 90 feature maps of size $(4\times3\times3)$. We spread each feature map as a structural feature vector of the corresponding brain region, and we can obtain the structural feature matrix $P \in R^{90\times36}$ of the brain region.

\paragraph{\textbf{Structure Profile.}}
It is still an open field of research on how to obtain the structural connectivity matrix $\hat{A}$ from the structural feature matrix $P$ of brain regions. A common approach is calculating the correlation, covariance, and mutual information between the structural feature vectors of brain regions\cite{deep}. One of the simplest methods is to make $\hat{A}=PP^T$ and then normalize $\hat{A}$. However, this approach relies excessively on structural features of brain regions and does not make full use of classification knowledge and structural network reconstruction knowledge. In the work of this paper, we use a more sophisticated parameterization method: we obtain the brain structural connectivity matrix $\hat{A}$ by learning a structural contour matrix $M$. The specific method is shown in Equation \ref{equation:eq1}.

\begin{equation}
    \hat{A}=exp(-|PMP^T|) \label{equation:eq1}
\end{equation}

Where $M\in R^{36\times36}$ is the learnable structural contour matrix and $exp(-|\cdot|)$ is the nonlinear mapping function.

We compute the empirical distribution $A$ of the structural connectivity matrix according to the AAL template, which is a symmetric matrix and all elements lie between $[0,1]$, indicating the relative connectivity strength between brain regions. To make the generated structural connectivity matrix $\hat{A}$ as close to the empirical distribution as possible, we use the nonlinear mapping function $exp(-|\cdot|)$ to restrict the elements to $[0,1]$. Moreover, according to the definition of structural brain network, $\hat{A}$ is a symmetric matrix. If we follow Eq.\ref{equation:eq1}, the resulting structural connection matrix may not be symmetric. Therefore, we further constrained $M$ such that $M=MM^T$ and obtained Eq.\ref{equation:eq2}.

\begin{equation}
    \hat{A}=exp(-|PMM^TP^T|) \label{equation:eq2}
\end{equation}

Although the formula of $\hat{A}$ has changed, the structural contour matrix $M$ to be learned remains unchanged. To avoid introducing bias, we initialize $M$ as a identity matrix in the training phase.

\subsection{Discriminator}
If a structural connection matrix $\hat{A} \sim P_{\hat{A}}$ generated by the generator and a structural connection matrix $A \sim P_{A}$ from the AAL template, then the role of the discriminator is to determine whether the input samples come from the distribution $P_{\hat{A}}$ or empirical distribution $P_{A}$ and to distinguish them maximally. The goal of the generator, on the other hand, is to maximize the learning of the empirical distribution $P_{A}$ and eventually reach $\hat{A} \sim P_{A}$, but this is only the ideal case. Combining the generator and discriminator to get a GAN for simultaneous training and introducing W-divergence(W-div) to measure the distance between the two distributions can solve the problems such as training difficulties and training instability. The loss functions of the discriminator and generator are shown in Eq. \ref{equation:eq3} and Eq. \ref{equation:eq4}, respectively.

\begin{equation}
    L_D=E_{\tilde{x} \sim P_{\hat{A}}}D(\tilde{x}) - E_{x \sim P_A}D(x) + E_{(\tilde{x},x) \sim (P_{\hat{A}},P_A)}k{||\nabla T||}^p
    \label{equation:eq3}
\end{equation}

\begin{equation}
    L_G= -E_{\tilde{x} \sim P_{\hat{A}}}D(\tilde{x})
    \label{equation:eq4}
\end{equation}

Where $\tilde{x},x$ denotes the sample input to discriminator D, $(\tilde{x},x) \sim (P_{\hat{A}},P_A)$ denotes the joint distribution of $P_{\hat{A}}$ and $P_A$, $\nabla T$ is the distance between the 2 distributions, $k,p$ is the hyperparameter.

\begin{equation}
    k{||\nabla T||}^p = \frac{k}{2} {[\sum{(\frac{\partial D(\tilde{x})}{\partial \tilde{x}})^2}]}^ \frac{p}{2} + \frac{k}{2} {[\sum{(\frac{\partial D(x)}{\partial x})^2}]}^ \frac{p}{2}
    \label{equation:eq5}
\end{equation}

We use two fully connected layers as the discriminator, with 8100 neurons in the input layer, ten neurons in the hidden layer, and one neuron in the output layer. The hidden layer uses $LeakeyReLU$ as the activation function while the output layer doesn't, and the output value is a scalar.

\subsection{Classifier}
\begin{figure}
	\centering
	\includegraphics[width=0.8\linewidth]{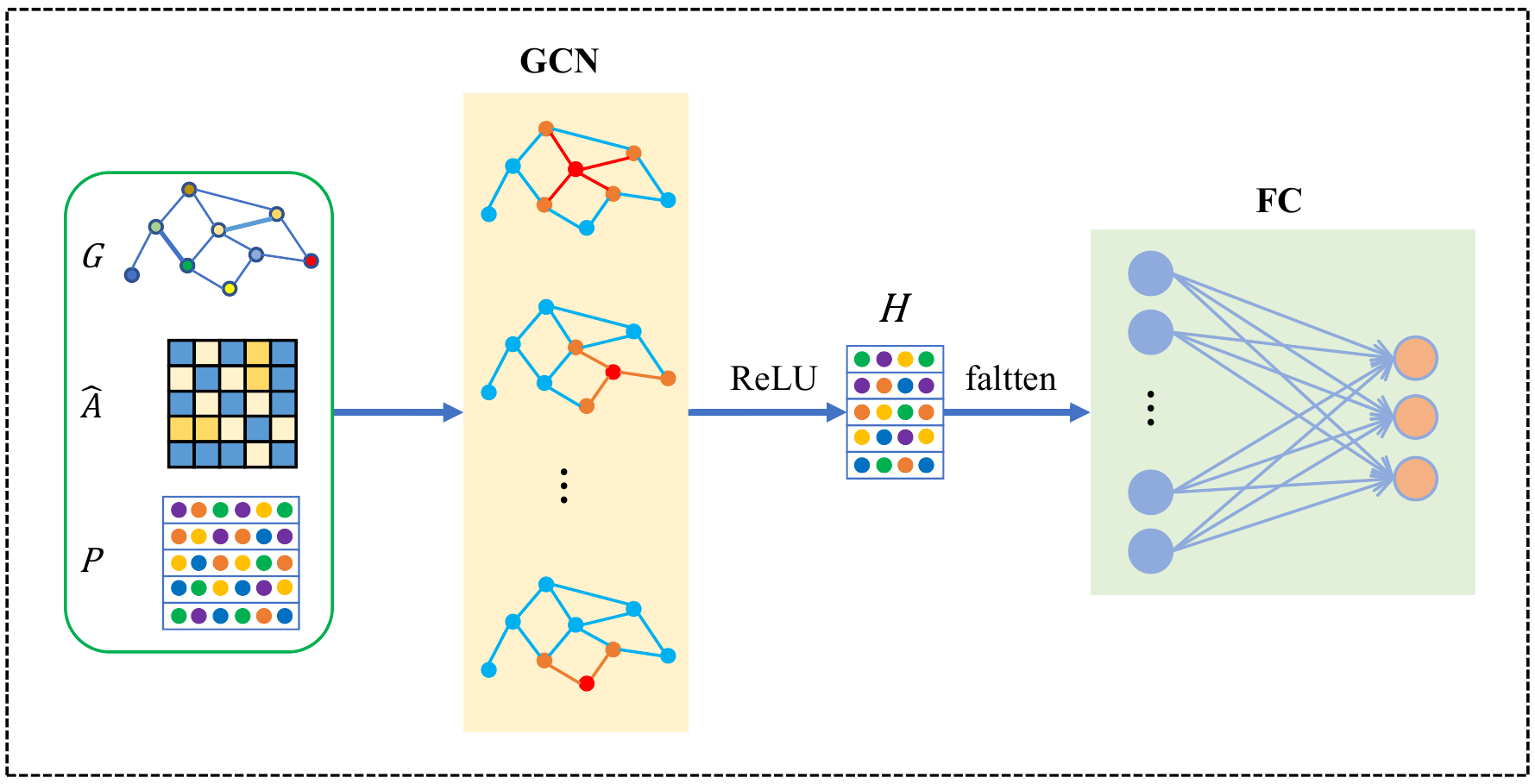}
	\caption{Framework of Classifier.}
	\label{fig:fig3}
\end{figure}

As shown in Figure \ref{fig:fig3}, the classifier consists of a GCN layer and a fully connected layer, where the GCN\cite{gcn} layer is used to fuse the structural features of each node's neighboring nodes, and the fully connected layer performs whole graph level classification based on the fused feature matrix. The classifier is used to classify EMCI, LMCI, and NC, but more importantly, to learn the structural brain network using the classification knowledge. Using Laplace transform on the structural connectivity matrix not only requires a higher computational cost but also may have an unknown effect on the results. Therefore, we did not use Laplace transform for the structural connection matrix when using GCN to fuse the feature information of different nodes in the structural brain network. The convolution process is shown in Eq. \ref{equation:eq6}.

\begin{equation}
	H= ReLU((\hat{A} + I)PW)
	\label{equation:eq6}
\end{equation}

Where $H$ denotes the structural feature matrix after convolution, $ReLU$ is the activation function of the convolution layer, $W \in R^{F_{in} \times F_{out}}$ is the weight matrix, $F_{in}=36, F_{out}=18$. Considering the structural characteristics of each node itself, we add an identity matrix $I$ to $\hat{A}$.

The number of input neurons of the fully connected layer is 3240, the number of output neurons is 3, and the activation function of the output layer is $ReLU$. We use the output results of the classifier and the class labels to calculate the cross-entropy loss as the loss of the classifier, as shown in Eq. \ref{equation:eq7}.

\begin{equation}
	L_C = - \frac{1}{N} \sum_{i=1}^N{p(y_i|x_i) log[q(\hat{y}_i|x_i)]}
	\label{equation:eq7}
\end{equation}

Where $N$ is the number of input samples, $p(y_i|x_i)$ denotes the actual distribution of sample labels, and $q(\hat{y}_i|x_i)$ denotes the distribution of labels at the output of the classifier.

\subsection{Trainning strategy}
We train the generator, discriminator, and classifier simultaneously, fixing the other two modules while training one of them. The discriminator is first trained by feeding the generated structural connection matrix $\hat{A}$ and the structural connection matrix $A$ from the AAL template into the discriminator and learning the parameters of the discriminator based on the discriminator loss $L_D$. The distance between distributions using the W-div metric does not cause the problem that the generator gradient disappears because the accuracy of the discriminator is too high, so we repeat the training of the discriminator five times during the iteration, which can speed up the convergence of the model. After training the discriminator five times, we train the classifier one time and update the classifier's parameters according to the classification loss $L_C$. Finally, we train the generator one time. In order to integrate the classification knowledge into the generation of the structural brain network, we calculate the weighted sum of the generator loss $L_G$ and the classification loss $L_C$ as the total loss of the generator, as shown in Eq. \ref{equation:eq8}, and then update the parameters of the generator using the $L_{G\_total}$.

\begin{equation}
    L_{G\_total}=L_G + \beta L_C \label{equation:eq8}
\end{equation}

Where $\beta$ is the weight coefficient, which is a hyperparameter that determines the relative importance of the classification loss $L_C$ in training the generator. In the early stage of model training, the pattern of the structural brain network generated by the generator is relatively simple, and the classification loss $L_C$ calculated using such a structure network is almost constant, which not only cannot guide the learning process of the classifier and generator but may even cause gradient cancellation. To prevent this from happening, we used linearly increasing weight coefficient $\beta$, and dynamically adjusted $\beta$ values according to the training process, as shown in Eq. \ref{equation:eq9}.

\begin{equation}
	\beta = \frac{\beta_{max} - \beta_{min}}{Epoches} \times epoch
	\label{equation:eq9}
\end{equation}

Where $\beta_{max}$ is the maximum value of the weight coefficient, $\beta_{min}$ is the minimum value of the weight coefficient, $Epoches$ is the maximum number of iterations, and $epoch$ is the current number of iterations.

\section{Experiments}
\subsection{Dataset}
The primary purpose of this paper is to investigate the differences in the structural brain network between EMCI, LMCI, and NC subjects, so we used only DTI as the dataset. DTI of 298 subjects were selected from the ADNI database for this study, including 87 NC subjects (42 males and 45 females, mean age 74.1 years, standard deviation 7.4), 135 EMCI patients (69 males and 66 females, mean age 75.8, standard deviation 6.4), 76 LMCI patients (35 males and 41 females, mean age 75.9 years, standard deviation 7.5).

For the DTI, we performed cranial stripping, resolution resampling, eddy current correction, and fiber tracking preprocessing operations using the PANDA toolbox. The voxel of the resampled 3-dimensional DTI was $109 \times 91 \times 91$. Then we divided the whole brain into 90 brain regions using the AAL template and counted the number of DTI-derived fibers between brain regions to calculate the structural connectivity matrix $A$. Because the number of fiber tracts varies widely between brain regions, we normalized $A$ to $[0,1]$ as the empirical distribution of the discriminator's input.

\subsection{Experimental settings}
For each subject, we obtained their DTI and structural connectivity matrix $A$ and added labels as input data for the model. We divided 298 subjects into a training set containing 250 subjects and a testing set containing 48 subjects (16 each for EMCI, LMCI, and NC subjects). We used batch training to train the model with $batch\_size=14$.

We detail the structure and parameters of the generator, discriminator, and classifier, as well as the training strategy of the model in Section 2. We train the three modules separately, thus defining 3 Adam optimizers with a learning rate of 0.005 for the generator, 0.001 for the discriminator, and 0.001 for the classifier. The hyperparameters in the discriminator loss $L_D$ are set to $k=2, p=6$, the joint loss $L_{G\_total}$ of the generator is taken to be $\beta_{max}=3$ and $\beta_{min}=0.1$, and the maximum number of iteration of the model is $Epoches=300$.

\subsection{Results}
\begin{figure}
	\centering
	\includegraphics[width=1\linewidth]{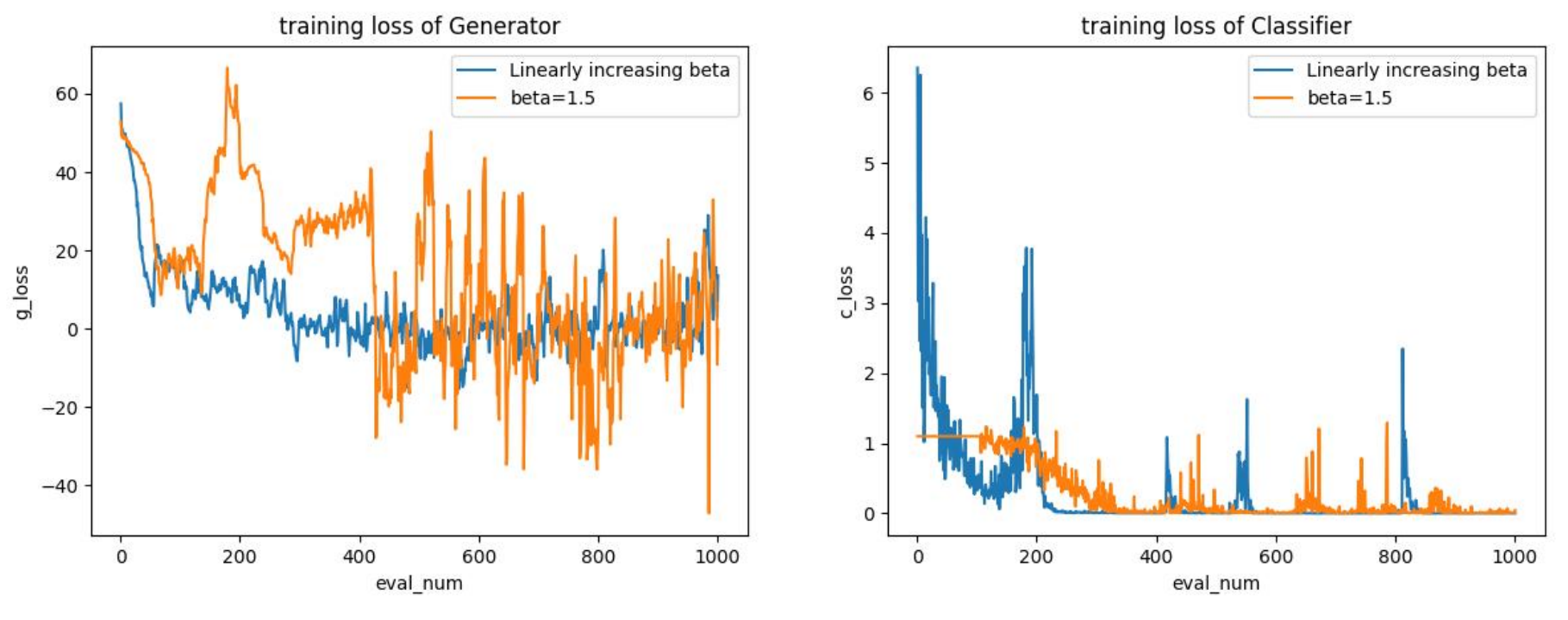}
	\caption{Generator loss and discriminator loss curves for 2 training strategies.}
	\label{fig:fig4}
\end{figure}

\paragraph{\textbf{Training process.}}
To demonstrate the effectiveness of the training strategy, we present in Figure \ref{fig:fig4} the training loss function curves using linearly increasing weight coefficient $\beta$ versus using fixed weight $\beta =2$. As can be seen from the figure, the training process of the generator is more stable using a linearly increasing weight coefficient, and the loss function of the classifier is no longer a horizontal line at the early stage of training.

\paragraph{\textbf{Classification Performance.}}
In the current study, we were more interested in patients in the MCI phase, so the subjects selected for the study were mainly NC, EMCI, and LMCI subjects, and these subjects were tri-classified. Currently, there is a paucity of research work using these three subjects for tri-classification. Therefore, in this section, we mainly show the test results of our two training strategies, including accuracy, precision, recall, and f1-score, as shown in Table \ref{table:tabel1}.

\begin{table}[]
	\centering
	\caption[]{Average detection performance of different training strategies.(\%)}
	\begin{tabular}{p{2.5cm}p{1.5cm}p{1.5cm}p{1.5cm}p{1.5cm}}
		\toprule
		& Accuracy & Precision & Recall & F1-score \\\midrule
		$\beta$=0.01    & 61.04    & 56.83     & 58.53  & 56.31    \\
		$\beta$=1.5     & 72.50    & 68.87     & 64.25  & 65.08    \\
		$\beta$=3       & 67.77    & 62.50     & 61.33  & 60.83    \\
		$\beta$-dynamic & \textbf{83.33}    & \textbf{88.58}     & \textbf{74.28}  & \textbf{76.75}   \\\bottomrule
		\label{table:tabel1}
	\end{tabular}
\end{table}

It can be seen from Table \ref{table:tabel1} that better classification results can be achieved using linearly increasing weight coefficient, with a classification accuracy of 83.3\%.

\paragraph{\textbf{Structural brain network generation.}}
Generating a structural brain network from DTI based on classification and reconstruction knowledge was our main goal in this study, and we tested the performance of the generator on 48 test samples. Our generator reconstructs the structural connectivity matrix of the brain well compared to the structural connectivity matrix from the AAL template. However, there are some differences in connectivity between local brain regions. This is because we introduced disease-related classification knowledge, and the generated structural connectivity matrix may better reflect patients' real brain structural connectivity. We hope to identify some brain regions that are highly correlated with MCI by analyzing the connectivity differences between these local brain regions.

\begin{figure}
	\centering
	\includegraphics[width=0.8\linewidth]{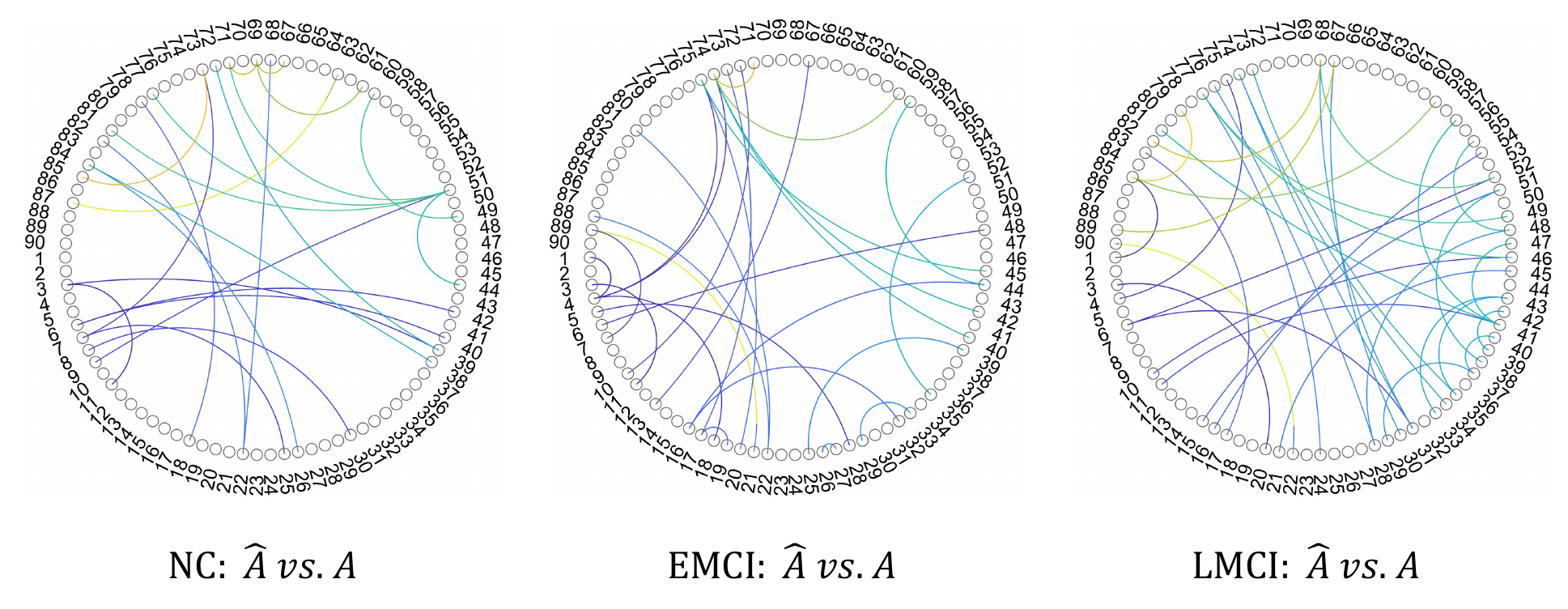}
	\caption{Changes in structural brain connectivity at different stages.}
	\label{fig:fig5}
\end{figure}

\paragraph{\textbf{Brain structural network connectivity analysis.}}
To analyze whether the structural connectivity matrix $\hat{A}$ generated by our proposed model was significantly changed compared to empirical distribution $A$, we tested $\hat{A}$ against $A$ using a paired-samples T-test at the significance level of 0.05. Dividing the testing set into three groups, EMCI, LMCI, and NC, with 16 individuals in each group, we used the generator to generate their structural connectivity matrix $\hat{A}$ and T-test it against the structural connectivity matrix $A$ from the AAL template, using a spin diagram to indicate those connections that have significant changes. As shown in Figure \ref{fig:fig5}, it can be found that the structural connectivity matrix $\hat{A}$ generated by our model has altered connectivity between many brain regions compared to empirical distribution $A$.

\begin{figure}
	\centering
	\includegraphics[width=1\linewidth]{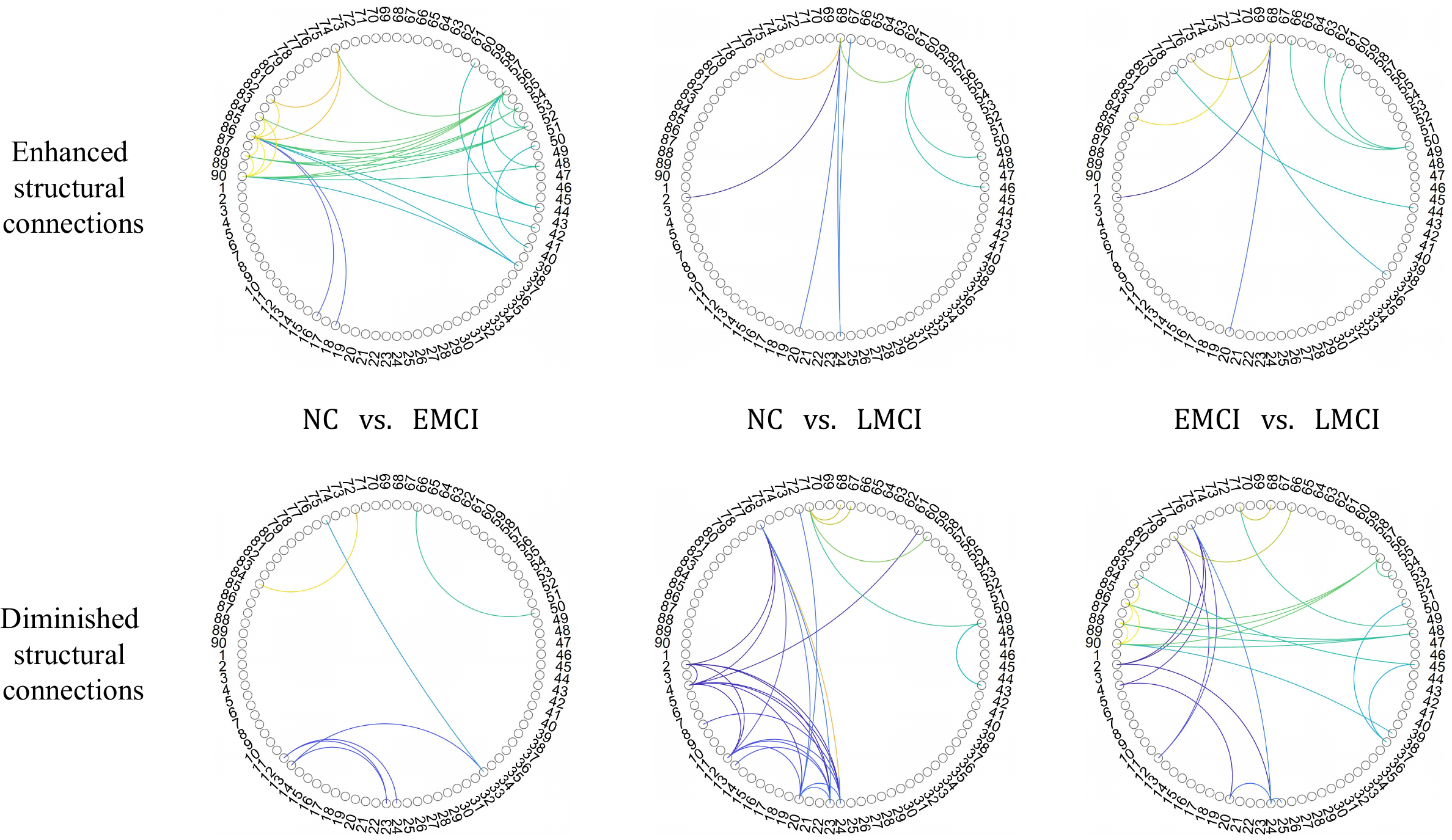}
	\caption{Spin diagram of connectivity change of SBGM generated connection matrix with empirical distribution A.}
	\label{fig:fig6}
\end{figure}

To further analyze the differences in the distribution of the brain structural connectivity matrix generated by our proposed model between different types of subjects, we compare the changes in brain connectivity in NC versus EMCI, NC versus LMCI, and EMCI versus LMCI patients, respectively, in Figure \ref{fig:fig6}. This change in connectivity between brain regions was more pronounced in LMCI patients compared with NC subjects. Similarly, between EMCI patients and LMCI patients, there was a significant decrease in connectivity between brain regions in LMCI patients. These changes and trends reveal a stepwise progression toward AD pathology in NC subjects: a gradual decrease in connectivity of the structural brain network.

\section{Conclusion}
In this paper, we develop a generative model for the direct generation of structural brain networks from DTI and diagnosing MCI patients. Our proposed model not only allows for more accurate classification of three subjects, EMCI, LMCI, and NC. More importantly,  it provides a method to generate structural brain networks directly from DTI, which is free from the limitations of many existing software templates and toolboxes, providing a prerequisite for us to study the changes in the structural brain networks of MCI patients. Experimental results on the ADNI dataset demonstrate that our proposed model achieves an accuracy of 83.4\% for the classification of three subjects, EMCI, LMCI, and NC. Moreover, we found that the brain structural connectivity of subjects gradually decreased from NC to EMCI to LMCI, which is consistent with some research findings in the field of neuroscience. In future research, we will focus on which brain regions with greater changes in structural connectivity of patients with AD as their disease worsens and extend our research work to the fusion of multimodal information.

\end{document}